\documentclass[twoside,leqno,twocolumn]{article}  
\usepackage{ltexpprt}

\usepackage{amssymb}
\usepackage{amsmath}
\usepackage{graphicx}
\usepackage{color}

\begin{document}

\newcommand{\quod}{\hfill $\blacksquare$ \bigbreak}
\newcommand{\reals}{I\!\!R}

\newcommand{\np}{\mbox{{\sc NP}}}
\newcommand{\sing}{\mbox{{\sc Sing}}}
\newcommand{\prob}{\mbox{Prob}}
\newcommand{\atm}{\mbox{{\sc ATM}}}
\newcommand{\hopn}{\hop_{\cN}}
\newcommand{\atmn}{\atm_{\cN}}
\newcommand{\cA}{{\cal A}}
\newcommand{\cO}{{\cal O}}
\newcommand{\cP}{{\cal P}}
\newcommand{\cC}{{\cal C}}
\newcommand{\C}{{\cal C}}
\newcommand{\cB}{{\cal B}}
\newcommand{\cG}{{\cal G}}
\newcommand{\cN}{{\cal N}}
\newcommand{\cU}{{\cal U}}
\newcommand{\cF}{{\cal F}}
\newcommand{\cT}{{\cal T}}
\newcommand{\hx}{\hat{x}}
\newcommand{\cS}{{\cal S}}

\newcommand{\MS}{\mathcal{S}}

\newcommand{\eps}{{\epsilon}}
\newcommand{\la}{{\lambda}}
\newcommand{\al}{{\alpha}}
\newcommand{\qed}{\hfill $\square$ \smallbreak}

\newcommand{\UDGI}{{\tt UDG1}}
\newcommand{\UDGII}{{\tt UDG2}}
\newcommand{\SYM}{{\tt SYM}}

\newcommand{\mode}{\mathop{mode}}
\newcommand{\true}{\mathop{true}}
\newcommand{\false}{\mathop{false}}
\def\rv{r_{\mathop{hist}}}
\def\rs{r(s_1)}
\def\rw{r_{\mathop{sim}}}
\newcommand{\inv}[1]{\overline{#1}}

\def\RV{\mbox{{\tt GraphRV}}}
\def\RVREC{\mbox{{\tt GraphRVREC}}}
\def\SIGMA{\mathop{SIGMA}}
\def\ROUTE{\mathop{ROUTE}}
\def\bbR{\mathbb{R}}
\def\bbN{\mathbb{N}}
\newcommand{\cR}{{\cal R}}
\newcommand{\con}{~^\smallfrown~}
\newcommand{\bbQ}{\mathbb{Q}}

\title{{\Large How to meet asynchronously (almost) everywhere}}
\author{
Jurek Czyzowicz\footnotemark[1]  \footnotemark[2]
\and
Arnaud Labourel \footnotemark[1] \footnotemark[3]\footnotemark[5]
\and
Andrzej Pelc \footnotemark[1] \footnotemark[4]
}
\date{ }
\maketitle
\def\thefootnote{\fnsymbol{footnote}}

\footnotetext[1]{ \noindent
D\'epartement d'informatique, Universit\'e du Qu\'ebec en Outaouais, Gatineau,
Qu\'ebec J8X 3X7, Canada.
E-mails: {\tt jurek@uqo.ca}, {\tt labourel.arnaud@gmail.com}, {\tt pelc@uqo.ca}  
}
\footnotetext[2]{ \noindent
Partially supported by NSERC discovery grant.
}
\footnotetext[5]{ \noindent
Supported by the ANR-project
  ``ALADDIN'', and the \'equipe-projet INRIA ``C\'EPAGE''.
}
\footnotetext[3]{ \noindent
This work was done during this author's stay at the
Universit\'e du Qu\'{e}bec en Outaouais as a postdoctoral fellow.  
} 
\footnotetext[4]{ \noindent  
Partially supported by NSERC discovery grant and     
by the Research Chair in Distributed Computing at the
Universit\'e du Qu\'{e}bec en Outaouais. 
}

\begin{abstract}\small\baselineskip=9pt
Two mobile agents (robots) with distinct labels have to meet in an arbitrary, possibly infinite,
unknown connected graph or in an unknown connected terrain in the plane. Agents are modeled as points, and the route
of each of them only depends on its label and on the unknown environment.
The actual walk of each agent also
depends on an asynchronous adversary that may arbitrarily vary the speed of the agent, stop it, 
or even move it back and forth, 
as long as the walk of the agent in each segment of its route is continuous, does not leave it and covers all of it.
Meeting in a graph means that both agents must be at the same time in some node or 
in some point inside an edge of the graph,
while meeting in a terrain means that both agents must be at the same time in some point of the terrain.  
Does there exist a deterministic algorithm that allows any two agents
to meet in any unknown environment in spite of this very powerfull adversary? 
We give  deterministic rendezvous algorithms for agents starting at arbitrary nodes of any anonymous connected graph
(finite or infinite) and for agents starting at any interior points with rational coordinates in any 
closed region of the plane with path-connected interior. 
While our algorithms work in a very general setting -- agents can, indeed, meet almost everywhere --
we show that none of the above few limitations imposed on the environment can be removed. On the other hand,
our algorithm also guarantees the following {\em approximate rendezvous} for agents starting at {\em arbitrary} interior
points of a terrain as above: agents will eventually get at an arbitrarily small positive distance from each other.
%\vspace*{1cm}
%\noindent
%{\bf keywords:} rendezvous, graph, asynchronous, deterministic
%\vspace*{4cm}
\end{abstract}

\section{Introduction}

{\bf The problem and the model.}
Two mobile agents (robots) modeled as points starting at different locations of an unknown environment have to meet.
This task is known in the literature as the rendezvous problem, and has been studied under two alternative scenarios.
Either the agents move in a network, modeled by an undirected connected graph (the {\em graph scenario}),
or they move in (some subset of) the plane (the {\em geometric scenario}). 

In this paper we study the {\em asynchronous} version of the rendezvous problem,
under both above scenarios: each agent designs its route and
an adversary controls the speed of each agent, can vary this speed, stop the agent, 
or even move it back and forth, 
as long as the walk of the agent in each segment of its route is continuous, does not leave it and covers all of it.
In the asynchronous version of the graph scenario,
meeting at a node may be impossible even in the two-node graph, as the adversary can  
desynchronize the agents and make them visit nodes at different times. Thus it is necessary
to relax the requirement
and allow agents to meet either in a node or inside an edge. Such a definition
of meeting is natural,
e.g., when agents are robots traveling in a labyrinth.
We consider an embedding of the
underlying graph in the
three-dimensional Euclidean space, with nodes of the graph being points of the
space and edges being
pairwise disjoint line segments joining them
(hence there are no edge crossings). Agents are modeled
as points moving inside this embedding. 

If nodes of the graph are labeled and the labeling is known, 
then agents can decide to meet at a predetermined node
and the rendezvous problem reduces to graph exploration. 
However, in many applications, when rendezvous
is needed in a network of unknown topology, such unique labeling of nodes may
be unavailable, 
or agents may be unable to perceive such labels, e.g., due to security reasons. 
Hence it is important to design rendezvous algorithms for
agents operating in {\em anonymous} graphs, i.e., graphs without unique labeling
of nodes.
It is important to note that the agents have to be able to {\em locally} distinguish ports at a
node: otherwise, an agent may even be unable to
visit all neighbors of a node of degree 3 (after visiting the second neighbor,
the agent cannot distinguish
the port leading to the first visited neighbor from that leading to the
unvisited one). Consequently, agents
initially located at two nodes of degree 3, might never be able to meet. 
This justifies a common 
assumption made in the literature: all ports at a node are locally labeled by distinct positive integers.
Degrees of nodes can be either finite or infinite.
No coherence between those local labelings is assumed. When an agent leaves a
node, it is aware of the port number
by which it leaves and when it enters a node, it is aware of the entry port
number. It can also verify, at each node, whether a given positive integer is a port number at this node.
Agents know neither the graph, nor the initial distance between them. They cannot mark the nodes or the edges in any way.
Rendezvous has to be accomplished regardless of local labelings of ports. 
Each agent terminates its walk at the time of meeting the other agent.

In the geometric scenario, we assume that the terrain in which agents operate
is a closed subset of the Euclidean plane, i.e., it contains limits of all converging sequences of points in it.
The {\em boundary} of the terrain is defined as the set of points having
arbitrarily close points
both in the terrain and outside of it. 
Since the terrain is closed, the boundary is included in it. 
All other points of the terrain are its {\em interior} points.  
Each agent can only distinguish if it is currently in an interior point of 
the terrain or in its boundary. Agents do not know the terrain in which they operate, 
they cannot ``see'' any vicinity of the currently visited point and they cannot leave any marks.
Agents are equipped with a compass and with the same unit of length.
Thus systems of coordinates of agents are aligned and the origin for each agent is at its starting point.
Again, an agent terminates
its walk at the time of meeting the other agent.

If agents are identical, i.e., they do not have distinct identifiers, and
execute the same algorithm, then
deterministic rendezvous is impossible, e.g.,  in the  ring (graph scenario)
or in the plane (geometric scenario): the adversary
will make the agents move always in the
same direction at the same speed, keeping them at the same distance at all times, thus they will never meet. 
Hence we assume that agents
have distinct identifiers, called labels, which are two different positive integers.
This is their only way to break symmetry.
We assume that each agent knows its own label but not the label of the other agent. 
This excludes, e.g.,  rendezvous strategies of the type ``waiting for mommy'', in which the agent with smaller label
remains idle and the other agent explores the graph or the terrain. We do not impose any restriction on the memory
of the agents: from the computational perspective they are viewed as Turing machines.
 
Two important notions used to describe movements of agents are the {\em route} of each agent and its {\em walk}.
Roughly speaking, each agent chooses the route {\em where} it moves and the adversary describes the walk on this 
route, deciding {\em how} the agent  moves. More precisely,  these notions are defined as follows.
The adversary initially places an agent with label $\ell$ at some node of the graph or at some point in the terrain.
Given this label and this starting point, the route is
chosen by the agent and is defined as follows. In the case of the graph, the agent chooses one of the available ports at the current node. After getting to the other end of the corresponding edge, the agent
chooses one of the available ports at this node, and so on, indefinitely (until rendezvous). 
The resulting route of the agent is the corresponding sequence of edges, 
which is a (not necessarily simple) path in the graph. 
The route in a terrain is a sequence $(S_1,S_2,\dots)$ of segments, where $S_i=[a_i, a_{i+1}]$, defined in stages as follows, given the agent's label and the starting point.
In stage $i$ the agent starts at point $a_i$, and $a_1$ is the starting point chosen by the adversary.
The agent chooses a direction $\alpha$ and distance $x$. 
If the segment of length $x$ in direction $\alpha$ starting in $v$ intersects the boundary
of the terrain at some distance $y\leq x$ from $v$, the agent becomes aware of it in the intersection point $w$ closest to $v$. In this case, the stage ends
at $w$ and in the next stage the agent chooses the reverse direction $\overline{\alpha}$ and the distance $y$ (that will cause it to return to $v$). If the segment of length $x$ in direction $\alpha$ starting in $v$ does not intersect the boundary of the terrain, the stage ends when the agent reaches point $u$ at distance $x$ from $v$ in direction $\alpha$.
Stages are repeated indefinitely (until rendezvous). 

We now describe the walk $f$ of an agent on its route. Let $R=(S_1,S_2,\dots)$ be the route of an agent. In the graph scenario this is a 
(not necessarily simple) infinite path in the (spatial embedding of) the graph, and in the geometric scenario it is an infinite polygonal line in the plane. Let $(t_1,t_2,\dots)$, where $t_1=0$, be an increasing sequence of reals, chosen by the adversary, that represent points in time. Let $f_i:[t_i,t_{i+1}]\rightarrow [a_i,a_{i+1}]$ be any continuous function, chosen by the adversary, such that $f_i(t_i)=a_i$ and $f_i(t_{i+1})=a_{i+1}$. For any $t\in [t_i,t_{i+1}]$, we define $f(t)=f_i(t)$. 
The interpretation of the walk $f$ is as follows: at time $t$ the agent
is at the point $f(t)$ of its route.  This general definition of the walk and the fact that it is constructed by the adversary
capture the asynchronous characteristics of the process. The movement of the agent can be
at arbitrary speed, the agent may sometimes stop or go back and forth, as long as the walk 
in each segment of the route is continuous and covers all of it. 

Notice that the power of the asynchronous adversary to produce any continuous walk on the 
routes determined by the agents implies the following significant difference with respect to the synchronous scenario. While in the latter scenario the relative movement of the agents depends only on their routes, in our setting, this movement is also controlled by the adversary.

Agents with routes $R_1$ and $R_2$ and with walks $f_1$ and $f_2$ meet at time $t$,
if points $f_1(t )$ and $f_2(t )$ are identical. A rendezvous is guaranteed for routes $R_1$ and $R_2$,
if the agents using these routes meet at some time $t$, regardless of the walks chosen by the adversary.
A rendezvous algorithm executed by agents in a graph or in a terrain produces routes of agents, given the label of each agent and its starting point.

It should be stressed that, while routes of agents are formally defined as infinite sequences of segments,
our results imply that in any instance of the rendezvous problem, meeting will occur at some finite time,
and thus each agent will compute only finitely many segments of its route.  As mentioned above, agents compute their
routes in stages, and given any walk chosen by the adversary, each stage is completed in finite time.
There is no stopping issue in
our solution: rendezvous always occurs at some stage for each of the agents and then both agents stop.
Another feature of our rendezvous algorithms is that in the choice of consecutive segments of its route an agent
does not use the knowledge of the walk to date. Thus the route depends only on the label of the agent, on the 
environment (graph or terrain), and on the starting point chosen by the adversary,  but not on its other actions.

{\bf Our results.}
We give two deterministic algorithms, the first for rendezvous in the graph scenario and the second in the geometric scenario. For the graph scenario, our algorithm accomplishes rendezvous in any connected countable\footnote{A graph is countable if the set of its nodes is countable, i.e., if there exists a one-to-one function from this set into the set of natural numbers.} (finite or infinite) graph, 
for arbitrary starting nodes. A consequence of this very general result is the positive answer to the following question from \cite{DGKKPV}: Is deterministic asynchronous rendezvous feasible in any finite connected graph without knowing any upper bound on its size?
(In \cite{DGKKPV} the authors presented a deterministic asynchronous rendezvous algorithm in arbitrary finite connected graphs with {\em known} upper bound on the size.)

For the geometric scenario, our algorithm accomplishes rendezvous
for agents starting at any interior points with  rational coordinates in any 
closed region of the plane with path-connected interior. (Recall that a subset $T$ of the plane is path-connected,
if for any points $u,v \in T$, there is a continuous function $h: [0,1] \longrightarrow P$, such that $P \subseteq T$
and $h(0)=u$, $h(1)=v$.)
On the other hand,
our algorithm guarantees the following {\em approximate rendezvous} for agents starting at {\em arbitrary} interior
points of a terrain as above: agents will eventually get at an arbitrarily small positive distance from each other.
This implies the perhaps surprising result that if agents have arbitrarily small positive visibility ranges
(rather than 0 visibility range as we assume) and they start in arbitrary points of the (empty) plane,
then they will see each other in finite time, 
regardless of the actions of the adversary. 

{\bf Discussion of limitations.}
While our algorithms work in a very general setting -- agents can, indeed, meet almost everywhere --
it turns out that none of the few limitations imposed on the environment can be removed. For the graph scenario,
the only limitation is connectivity of the graph. It is clear that rendezvous in disconnected graphs is impossible, 
if the agents start in different connected components. For the geometric scenario, let us review the limitations
one by one. First, we assume that the terrain is closed.  This assumption cannot be entirely removed for the following
technical reason.
Consider the construction of a route in an open disc. An agent starting at any point, that chooses in the first
stage an arbitrary direction
and a sufficiently large distance, at some point would have to leave the disc. Since it does not see anything in its vicinity,
it cannot know where the boundary is before hitting it, and it cannot hit it, as it is not allowed to leave the terrain.
It follows that the agent could not construct further segments of its route. The second assumption is that agents start
at interior points of the terrain. This assumption cannot be removed either. Indeed, suppose that the terrain is a closed disc
with a semi-circle attached to it.
This is a closed subset of the plane with nonempty path-connected interior. Suppose that one agent starts in the disc and the other 
at the end of the semi-circle. 
Since agents need to move along polygonal lines, the second agent could
not move at all and the first one cannot reach it. Our next assumption is that the interior of the terrain is path-connected.
To show that this assumption cannot be removed,
consider two disjoint closed discs joined by an arc of a circle. This terrain is closed and path-connected, but if each agent
starts inside a different disc, again they cannot meet, because agents need to move along polygonal lines, and hence cannot
traverse the joining arc.
The final assumption is that the starting points of the agents have rational coordinates.
 In Section 4 we prove that if the agents start in {\em arbitrary} points, then rendezvous cannot  be guaranteed
 even in the plane. We show, however, that for arbitrary starting points approximate rendezvous is guaranteed.
 
 {\bf Related work.}
The rendezvous problem was first described in \cite{schelling60}. A detailed discussion of the large 
literature on rendezvous can be found in the excellent book
\cite{alpern02b}. Most of the results in this domain can be divided
into two classes: those 
considering the geometric scenario (rendezvous in the line, see, e.g., 
\cite{baston98,baston01,gal99},
or in the plane, see, e.g., \cite{anderson98a,anderson98b}), and those
discussing rendezvous in graphs,
e.g., \cite{alpern95a, alpern99}. Some of the authors, e.g.,
\cite{alpern95a,alpern02a,anderson90,baston98,israeli} consider
the probabilistic scenario where inputs and/or rendezvous strategies are random. 
Randomized rendezvous strategies use random walks in
graphs, which
were thoroughly investigated and applied also to other problems, such as graph traversing
\cite{akllr}, on-line algorithms
\cite{cdrs} and estimating volumes of convex bodies \cite{dfk}.
A generalization of the rendezvous
problem is that of gathering \cite{fpsw,israeli,KKN,KMP,lim96,thomas92}, when more than
2 agents have to meet in one location.

If graphs are unlabeled, deterministic rendezvous requires breaking symmetry, which can be accomplished either
by allowing marking nodes or by labeling the agents.
Deterministic rendezvous with anonymous agents working in unlabeled graphs but 
equipped with tokens used to mark nodes was considered e.g., in~\cite{KKSS}.
In~\cite{YY} the authors studied gathering many agents
with unique labels. In \cite{DFKP,KM,TZ} deterministic rendezvous in graphs with
labeled agents was considered.
However, in all the above papers, the synchronous setting was assumed.
Asynchronous gathering under geometric
scenarios has been studied, e.g., in \cite{CFPS,fpsw,Pr} in different models than ours:
agents could not remember past events, but  
they were assumed to have at least partial visibility of the scene.
The first paper
to consider deterministic asynchronous rendezvous in graphs was \cite{DGKKPV}.
The authors concentrated on complexity of rendezvous in simple graphs, such as the ring
and the infinite line. They also showed feasibility of  deterministic asynchronous rendezvous in arbitrary finite
connected graphs with {\em known} upper bound on the size. Further improvements of the above results 
for the infinite line were proposed in~\cite{Sta09}. Gathering many robots in a graph, under a  different 
asynchronous model and assuming that the whole graph is seen by each robot, has been studied in \cite{KKN,KMP}.

\section{Preliminary notions and results}

A fundamental notion on which our algorithms are based is that of a {\em tunnel}.
Consider any graph $G$ and two routes $R_1$ and $R_2$ starting at nodes $v$ and $w$, respectively.
We say that these routes form a tunnel, if there exists a prefix $[e_1,e_2,\dots,e_n]$ of route $R_1$ 
and a prefix $[e_n, e_{n-1},\dots, e_1]$
of route $R_2$, for some edges $e_i$ in the graph, such that $e_i=\{v_i,v_{i+1}\}$, where $v_1=v$ and $v_{n+1}=w$. 
Intuitively, the route $R_1$ has a prefix $P$ ending at $w$ and the
route $R_2$ has a prefix which is the reverse of  $P$, ending at $v$.
By a slight abuse of terminology we will also say that prefixes $[e_1,e_2,\dots,e_n]$ and 
$[e_n, e_{n-1},\dots, e_1]$ form a tunnel.

\begin{proposition}\label{tunnel}
If routes $R_1$ and $R_2$ form a tunnel, then they guarantee rendezvous.
\end{proposition}

\begin{proof}
Consider an embedding of the graph $G$ in the three-dimensional Euclidean space, with nodes of the graph being points of the space and edges being pairwise disjoint line segments joining them. 
Consider routes $R_1$ and $R_2$ starting at nodes $v$ and $w$, respectively. 
Let agent $a_i$ execute route $R_i$. Let $P$ be the polygonal line joining $v$ with $w$, corresponding to the prefixes of the routes, given by the tunnel.  Let $D$ be its length defined as the sum of lengths of
edges in the corresponding prefixes of the routes. (For non-simple paths in the graph, the same edge is counted many times.)
%$P$ is a continuous image of a closed interval $[a,b]$.
%Let $\phi: [a,b] \longrightarrow P$ be a continuous function, 
%such that $\phi(a)=v$ and $\phi(b)=w$.
Consider any walks $f_1$ on $R_1$ and $f_2$ on $R_2$.
Let $t'$ be the first moment when an agent leaves its starting point and let $t''$ be the moment when an agent gets to the end of  $P$ other than its starting point. For any $t\in[t', t'']$, let $d_1(t)$ be the distance of agent $a_1$ from its starting point $v$ at time $t$, counted on the route $R_1$, and let $d_2(t)$ be the distance of agent $a_2$ from its target point $v$ at time $t$, counted on the route $R_2$. Let $\delta(t)=d_2(t)-d_1(t)$. We have $\delta(t')=D$ and $\delta(t'')=d \leq 0$. 
The function $\delta$ is thus a continuous function from the interval $[t', t'']$ onto some interval $[d',D]$, where $d'\leq d$, in view of the continuity of walks $f_1$ and $f_2$. Since 0 belongs to the interval  $[d',D]$, there must exist a moment $t$ in the interval  $[t', t'']$, for which $\delta(t)=0$. For this point we have  $f_1(t)=f_2(t)$, and the rendezvous occurs.
\end{proof}

We now recall some basic facts from set theory, that will be used in further considerations.
\begin{proposition}\label{sets}
The set of rational numbers and the set of positive rational numbers are countable.
The cartesian product of two countable sets is countable.
The set of all finite sequences with terms in a countable set is countable.
\end{proposition}

\section{Rendezvous in the graph scenario}

Let $G=(V,E)$ be the connected graph in which the rendezvous must be performed.
Let $\cS_n$ be the set of sequences of $n$ positive integers. 
Let $\cP=\{(i,j,s',s'')\mid i,j\in \bbN, i< j \mbox{ and } \exists n \mbox{ s.t. }s',s''\in \cS_n\}$. 
There exists a bijection from the set of positive integers onto $\cP$, in view of Proposition~\ref{sets}. 
Let $(\varphi_1,\varphi_2,\ldots)$ be a fixed enumeration of $\cP$. 
All agents have to agree on the same enumeration. It is easy to produce a formula computing $\phi _k$ for any $k$.
This formula is included in the rendezvous algorithm.
For a finite path $r$ in $G$, 
we denote by $\inv{r}$ the path with the same edges as in $r$, but in the reverse order. 
Remark that $r$ and $\inv{r}$ form a tunnel.

We first give a high-level idea of the algorithm referring to lines of the pseudo code given below.
We ``force'' the routes of any two agents to form a tunnel
for every possible combination of starting nodes and labels of the two agents. 
By Proposition~\ref{tunnel}, this suffices to guarantee rendezvous. 
Any starting configuration of robot $i$ placed at node $v$ and robot
$j$ placed at node $w$ by the adversary corresponds to a quadruple $(i,j,s',s'')$ 
where $s'$ is a sequence of ports inducing a path from $v$ to $w$ and $s''$ is a 
sequence of ports inducing the reverse path from $w$ to $v$. 

Each agent constructs its route in phases. At the beginning and at the end of each phase 
the agent is in its starting node. At phase $k$ the previously constructed initial 
part of the route $\rv$ is extended while the agent processes quadruple $\varphi_k$ 
(some of the extensions are null). This extension guarantees that the routes of agents 
of the corresponding  starting configuration will form a tunnel. When agent with 
label $l$ processes quadruple $\varphi_k=(i,j,s',s'')$ nothing happens if $l\neq i$ 
and $l\neq j$ (line 4). 
If $l=i$, agent $i$ tries to extend its route to guarantee rendezvous with agent $j$
under the hypothesis that a path from  $v$ to $w$ corresponds to the sequence $s'$ of ports and the reverse
path corresponds to the sequence $s''$. For this to happen, 
the agent first tries to follow the path $r(s')$ 
induced by the sequence $s'$ of ports (lines 8-11). This attempt is considered 
successful if the following conditions are satisfied:

$\bullet$ at consecutive nodes of the traversed path, ports with numbers from the sequence $s'$ are available,

$\bullet$ the reverse path corresponds to the sequence $s''$ of ports. 

When the attempt is successful (the condition of line 13 is satisfied) the agent is 
at node $w$ and it simulates the first $k-1$ phases of the execution of the algorithm 
by agent with label $j$ starting from $w$.  
The effect of this simulation is the path $\rw$.
Upon completion of this part, agent with label $i$ returns to $w$. 
Now the agent is able to further extend its path to form a tunnel with the route of 
agent $j$ (line 16). Finally, whether the attempt to follow the path $r(s')$ is 
successful or not, the agent with label $i$ backtracks to $v$ (line 17). If $l=j$, 
the above actions are performed with the roles of $i$ and $j$ reversed and the role of $s'$ and $s''$ reversed. 

Algorithm $\RV$ calls the recursive function $\RVREC$. This function is called in two 
different modes controlled by the boolean $\mode$. In the ``main'' mode $(\mode=\true)$ the function is executed indefinitely, until
rendezvous. In the ``simulation'' mode $(\mode=\false)$, the function is executed for all values up to a given $p$, 
or until rendezvous, whichever comes first. The symbol $\con$ denotes the concatenation of sequences.

\noindent\textbf{Algorithm $\RV$}\\
INPUT: A starting node $v\in V$ and a label $l$ of the agent.\\
OUTPUT: A rendezvous route $r$.

$\RVREC(v, l,0,\true)$;
\begin{tabbing}
{\bf function} \\
$\RVREC(\mbox{node } v, \mbox{label } l,\mbox{integer }p, \mbox{boolean }\mode)$\\
1~\= $k:=1; r:=\lambda;$\\
2 \> {\bf wh}\={\bf ile}~not rendezvous and $(k\leq p \mbox{ or } \mode)$ {\bf do}\\
3 \>\> let $\varphi_k=(i,j,s',s''); \rv:= r$;\\
4 \>\> {\bf if}~$l$\=~$=i$ or $l=j$ {\bf then}\\
5 \>\>\> {\bf if}~\=$l=i$ {\bf then} $s_1:=s'; s_2:=s''; l':=j;$\\
6 \>\>\> {\bf els}\={\bf e}~$s_1:=s''; s_2:=s'; l':=i;$\\
7 \>\>\> let $s_1=(p_1,\ldots,p_n); m:=1; \rs:=\lambda$;\\
8 \>\>\> {\bf while} $m\leq n$ and $p_m$ is a port {\bf do}\\
9 \>\>\>\> $\rs$\=$:=\rs\con(e_m)$\\ 
 \>\>\>\>\>where $e_m$ corresponds to port $p_m$;\\ 
10 \>\>\>\> let $a_m$ be the port corresponding\\
 \>\>\>\>\> to $e_m$ at its other endpoint;\\
11 \>\>\>\> $m:=m+1;$\\
12 \>\>\> $r:=r\con\rs;$\\
13 \>\>\> {\bf if}~$s_2=(a_n,\ldots,a_1)$ {\bf then} \\
14 \>\>\>\> let $w$ be the current node;\\
15 \>\>\>\> $\rw:=\RVREC(w,l',k-1,\false);$\\
16 \>\>\>\> $r:={r}\con{\rw}\con{\inv{\rs}}\con$\\
 \>\>\>\>\> ${\inv{\rv}}\con{\rs}\con{\inv{\rw}};$\\
17 \>\>\> $r:=r\con\inv{\rs};$\\
18 \>\> $k=k+1;$\\
19 \= \textbf{return} $r$ 
\end{tabbing}

\begin{figure}[h]
	\begin{center}
	\includegraphics{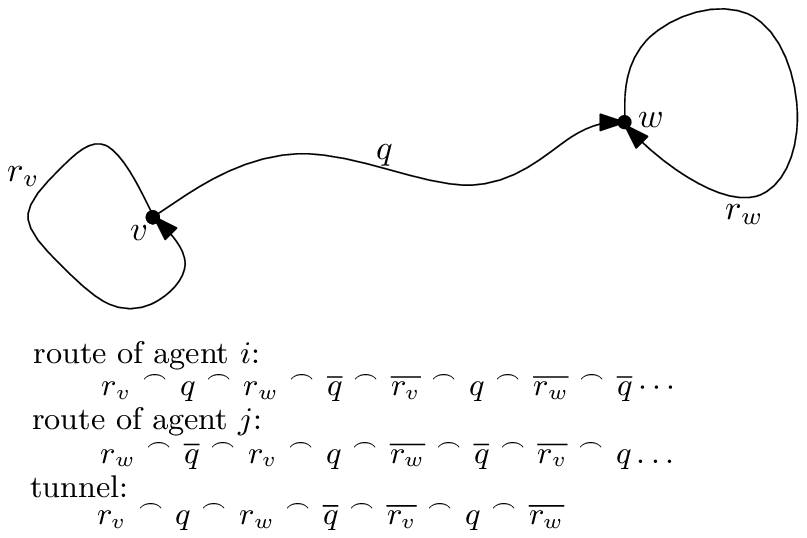}
	\caption{Tunnel between the routes of two agents}
	\label{fig:tunnel}
	\end{center}
\end{figure}

\begin{theorem}\label{th:correctness}
Algorithm $\RV$ guarantees asynchronous rendezvous for arbitrary two agents starting from any nodes of an arbitrary connected graph.
\end{theorem}

\begin{proof}
Let $v$ and $i$ (resp. $w$ and $j$) be the starting node and the label of the first agent (resp. the second agent). 
There exists a path $q$ linking $v$ to $w$, since the graph $G$ is connected. Let 
$s'$ (resp. $s''$) be the finite sequence of ports corresponding to the path $q$ (resp. the path 
$\inv{q}$). In view of Proposition~\ref{tunnel}, it suffices to prove that the routes 
of the two agents form a tunnel. We show that, after the phase corresponding to the quadruple 
$\varphi_k=(i,j,s',s'')$ during the execution of Algorithm $\RV$ for agents $i$ and $j$, the routes 
of the two agents form a tunnel. Observe that this phase eventually occurs during any execution of $\RVREC$, 
since all recursive calls of any phase $k'< k$ are done with a parameter $p$ strictly smaller than $k'$. 
Thus all these phases are completed in finite time.  

First, we show by induction that, at the beginning of phase $k$ of any execution of Algorithm
$\RV$, each agent is at its starting node. This is clearly true for $k=1$. Assume   
that the property holds for $k-1$. It follows that during the execution of the phase $k-1$, 
the paths $\rv$ and $\rw$ are cycles. Hence, after the execution of line 16, the agent ends in
node $w$. After the execution of line 17, the agent returns to the starting 
node $v$ of the phase $k-1$. So, the agent starts phase $k$ in the same node, and the property is true for all $k$. 

Let $r_{v}$ (resp. $r_{w}$) be the output of the execution of the first $k-1$ phases of Algorithm $\RV$
for agent $i$ (resp. $j$) starting in node $v$ (resp. $w$). At the beginning of phase $k$, the portion 
of the route constructed by agent $i$ is $r_v$. After the execution of line 12, the portion 
of the route constructed by agent $i$ is $r_v\con q$, since the agent has started the phase in node $v$. 
The path $\rw$ computed by the recursive call of $\RVREC$ is equal to $r_w$. It follows that at the end of phase $k$, the portion of the route constructed by agent $i$ is 
$\rho=r_{v}\con{q}\con{r_{w}}\con{\inv{q}}\con{\inv{r_{v}}}\con{q}\con{\inv{r_{w}}}\con{\inv{q}}$. 
Similarly, at the end of phase $k$, the portion of the route constructed by agent $j$ is 
$\rho'=r_{w}\con{\inv{q}}\con{r_{v}}\con{q}\con{\inv{r_{w}}}\con{\inv{q}}\con{\inv{r_{v}}}\con{q}$. 
By construction, the part $r_{v}\con{q}\con{r_{w}}\con{\inv{q}}\con{\inv{r_{v}}}\con{q}\con{\inv{r_{w}}}$ 
of $\rho$ and the part $r_{w}\con{\inv{q}}\con{r_{v}}\con{q}\con{\inv{r_{w}}}\con{\inv{q}}\con{\inv{r_{v}}}$ 
of $\rho'$ form a tunnel
(see Fig~\ref{fig:tunnel}).
\end{proof}

\section{Rendezvous in the geometric scenario}

In this section we consider the problem of rendezvous in a terrain included in the Euclidean plane.
As announced in the introduction, we restrict attention to
closed subsets of the plane whose interior is
path-connected. We observed that these restrictions cannot be removed.

Fix a system of coordinates $\Sigma$ with the $y$-axis pointing to North (shown by compasses of the agents) and with the unit
of length equal to that of the agents. 
Points with rational coordinates in $\Sigma$ will be called {\em rational}.
A polygonal line all of whose vertices, including extremities, are rational
will be called a {\em rational line}. For any point $u$, let $\Sigma _u$ be the shift of the system $\Sigma$
with origin at point $u$. 

\begin{lemma}\label{lm:rational-route}
In any path-connected, open subset $S$ of the plane and for any rational
points $u, v \in S$, there exists a rational polygonal line included in $S$, with
extremities $u$ and $v$, all of whose vertices are rational.
\end{lemma}

\begin{proof}
By path-connectivity of $S$,
there exists a path $p$ included in $S$ with extremities $u$ and $v$,
which is  a continuous image of the interval $[0,1]$. Let $d$ be the
distance from $p$ to $cS$ - the complement of $S$. Since $p \cap cS =
\emptyset$ and both $p$ and $cS$ are closed sets, we have $d>0$.
Partition the plane into squares of side length at most $d/2$ with rational vertices. 
Let $Q$ be the set of squares intersecting $p$. Since $p$ is a bounded set, $Q$ is
finite. Consider the graph $G_p$ with node set $Q$, such that $x,y \in
Q$ are adjacent if $p$ contains a point belonging to a common boundary
of $x$ and $y$. Since $G_p$ is connected, there exists a path $(x_1,\ldots,x_k)$
in $G_p$ linking squares $x_1$ containing $u$ and $x_k$ containing $v$. Let $p^*$
be the polygonal path $(\overline{uw_1},\overline{w_1w_2},\ldots,\overline{w_{k-1}w_k},\overline{w_kv})$,
where $w_i$ is the center of square $x_i$, for all $i=1,\ldots,k$. The path $p^*$ is
rational and contained in the union of squares from $Q$. 
Since each point of $p^*$ is at distance at most $d\sqrt{2}/2$
from some point of $p$, we have $p^* \cap cS = \emptyset$, hence $p^*$
is included in $S$.
\end{proof}

We define the following graph $G_T=(V, E)$, for a 
given terrain $T$.  
The set of nodes $V$ is the union of two disjoint
subsets $V_1, V_2$. The set $V_1$ is the set of all interior, rational
points of $T$ and the set $V_2$ is defined below. 

For each pair of
points $p_1, p_2$, such that $p_1 \in V_1$ and $p_2$ is any
rational point of the plane, we consider the segment
$s=\overline{p_1p_2}$. If $s$ does not intersect the boundary of $T$,
then  $p_2$ must belong to $V_1$ and we add the edge $\{p_1,p_2\}$ to
$E$. If $s$ intersects the boundary of $T$, we add a new node $v$ to
$V_2$ and we add the edge $\{p_1,v\}$ to $E$. 
Note that such a node $v$
is always added to $V_2$ when point $p_2$ is on the boundary of $T$ or
outside of $T$ (and it may or may not be added to $V_2$ when $p_2$ is in
the interior of $T$). 
Since each node in $V_2$ corresponds to a pair of
rational points $p_1, p_2$, there is a countable number of
nodes in $V_2$, each of them having degree 1. The unique port at any node
in $V_2$ has number 1 and ports at any node in $V_1$ are defined as follows.
Let $(z_1,z_2,\dots)$ be any fixed enumeration of all pairs of rational numbers.
Let $z_i=(q_1,q_2)$. Let $u$ be the point in the plane with coordinates $(q_1,q_2)$
in the system $\Sigma _p$. The port at $p$ corresponding 
to edge $\{p,u\}$ has number $i$.

{\bf Algorithm} {\tt GeometricRV}

The algorithm is a direct application of Algorithm {\tt GraphRV}
to the graph $G_T$. The agent operating in an
unknown terrain $T$ designs a route in the corresponding unknown graph
$G_T$ as follows. When the agent chooses a port at a node $p \in V_1$,
this edge corresponds to some rational point $q_i$ in the plane that the
agent tries to reach from $p$. Two cases may occur. 
The agent either walks in the interior of $T$ until reaching $q_i$, which corresponds to the
traversal of an edge between two nodes of $V_1$, or it hits the boundary
of $T$, which corresponds to a visit of a node $p'\in V_2$. At a node $p' \in V_2$
there is no choice of port, since its degree is 1. The agent takes the unique port which
leads to the (already visited) node $p \in V_1$. The resulting
route is a sequence of segments joining rational interior points of the terrain $T$ and
of pairs of consecutive segments $(\overline{vb},\overline{bv})$, where
$v$ is a rational interior point of $T$ and $b$ is a point on the
boundary of $T$. 

Since by Lemma \ref{lm:rational-route} graph $G_T$ is
connected, rendezvous is guaranteed in the graph $G_T$,
which implies rendezvous in $T$.

\begin{theorem}\label{GeometricRV}
Algorithm {\tt GeometricRV} guarantees asynchronous
rendezvous for arbitrary two agents starting from arbitrary rational interior
points of any closed terrain $T$ with path-connected interior.
\end{theorem}
  
Theorem \ref{GeometricRV} should be contrasted with the following negative result showing that the restriction on the starting
points of the agents cannot be removed, even for rendezvous in the (empty) plane.

\begin{proposition}\label{neg}
There is no algorithm that guarantees asynchronous rendezvous of arbitrary agents starting from arbitrary points in the plane.
\end{proposition}

\begin{proof}
Consider the agent with label $\ell$ operating in the empty plane. Since the terrain is fixed, the route of this agent
depends only on the starting point. Let $R=(e_1,e_2,...)$ be the route of the agent with label 1 starting at a fixed point $v$.
Consider the route $R_2(w)$ of the agent with label 2 starting at point $w$. Since there are no boundary points in the terrain,
for any starting points $w'$ and $w''$, route $R_2(w'')$ is a parallel shift of route $R_2(w')$  by the vector $(w',w'')$.
Both of them are polygonal lines.

We will say that two routes are {\em almost disjoint} if all vertices of each of them are outside the other route.
Observe that if, for some starting point $w$, route $R_2(w)$ of agent 2 is almost disjoint from route $R$ of agent 1,
then the adversary can avoid rendezvous of agents 1 and 2 following these routes, by first moving agent 1 to the end of the
first segment of its route, then moving agent 2 to the end of the
first segment of its route and so on, alternating traversals of agents on consecutive segments of their routes.
Hence, in order to prove our result, it is enough to show the existence of a point $w^*$,
such that route $R_2(w^*)$ is almost disjoint from route $R$.

Let $(f_1,f_2,...)$ be the sequence of vectors corresponding to consecutive segments of the route $R_2(w)$, for any starting point $w$.
For any fixed point $w$, denote by $f_j(w)$ the segment corresponding to vector $f_j$ on route $R_2(w)$. Let $(p_1,p_2,\dots)$
be any sequence that orders all couples $(e_i,f_j)$, for all positive integers $i,j$.
For any $k$, let $p_k=(e_{i_k},f_{j_k})$ 
We will construct by induction a descending sequence
of closed discs $(D_1,D_2,\dots)$ of positive radii, satisfying the following invariant.  
For all points $w \in D_k$, both endpoints of the segment $e_{i_k}$ are outside of the segment $f_{j_k}(w)$
and both endpoints of the segment $f_{j_k}(w)$ are outside of the segment $e_{i_k}$.
Suppose that the invariant is satisfied for $k-1$ (for $k=1$ we may take as $D_0$ any disc with radius 1).
The set of points $w$ inside disc $D_{k-1}$ that may possibly violate the invariant for $k$ is contained in the union of four segments:
two segments parallel to $f_{j_k}$ and two segments parallel to $e_{i_k}$. There exists a closed disc of positive radius contained
in $D_{k-1}$ which is disjoint from those four segments. Let $D_k$ be such a disc.
Thus the invariant is satisfied for $k$. This completes the construction by induction.

The intersection of all discs $D_k$ is non-empty. Let $w^*$ be a point in this intersection.
Since $(p_1,p_2,\dots)$ enumerated all couples $(e_i,f_j)$, it follows that route $R_2(w^*)$ is almost disjoint from route $R$,
and hence agents 1 and 2 starting at points $v$ and $w^*$, respectively, do not meet for some walks chosen by the adversary.
\end{proof}

While Proposition \ref{neg} shows that rendezvous of agents starting from arbitrary points is impossible, it
turns out that a slightly easier task can be accomplished in this setting. For any $\epsilon >0$, we say that 
routes $R_1$ and $R_2$ of agents guarantee $\epsilon$-{\em approximate rendezvous}, if at some point $t$ of time, 
the agents get at distance at most $\epsilon$ from each other,
regardless of the walks chosen by the adversary.   

\begin{theorem}\label{approx}
Algorithm {\tt GeometricRV} guarantees $\epsilon$-approximate rendezvous for any $\epsilon >0$, 
for arbitrary agents 
starting from arbitrary interior points of any closed terrain $T$ with path-connected interior.

\end{theorem}

\begin{proof}
Fix an $\epsilon >0$.
Consider any two agents starting at interior points $v$ and $w$ of the terrain $T$.
Let $\rho >0$ be the distance from $w$ to the boundary of $T$. Choose a point $w'$ with rational coordinates 
in $\Sigma _v$, in the interior of the disc $D$
of radius $\rho/2$ centered at $w$. By Lemma \ref{lm:rational-route}
(applied to the system $\Sigma _v$ instead of $\Sigma$), there exists a rational polygonal line $P$ 
in the system $\Sigma _v$,
included in the interior of $T$ and joining $v$ and $w'$.
Let $d>0$ be the distance between $P$ and the boundary of $T$. Let $r=\min (\rho/2,d/2,\epsilon)$.
Choose a point $w''$ with rational coordinates in the system  $\Sigma _v$, at distance less than $r$ from $w$.
Let $P_v$ be the polygonal line $P$ extended by the segment $\overline{w'w''}$. Note that $P_v$ is at distance at least $r$
from the boundary of $T$. Indeed, if $x \in P$, then the distance from $x$ to the boundary is at least $d>r$,
and if $x \in \overline{w'w''}$, then the  distance from $x$ to the boundary is at least $\rho/2 \geq r$, as the entire 
segment $\overline{w'w''}$ is included in disc $D$.

Let $\Phi$ be the translation by the vector $(w''w)$ and let $P_w$ be the image of $P_v$ with respect to $\Phi$.
The point $w$ is one of the extremities of $P_w$. Since the distance from $w''$ to $w$ is less than $r$, 
the entire polygonal line $P_w$ is in the interior of $T$. The polygonal line $P_w$ is rational in the system $\Sigma _w$.

Let $f$ be any walk of the first agent on any route with the initial part $P_v$ starting at $v$,
and let $g$ be any walk of the second agent on any route with the initial part $P_w$ starting at $w$.
Consider the composition $g^*= \Phi ^{-1} \circ g$. Hence $g^*$ is a walk on a route with initial
part $P_v$, starting at $w''$. By Theorem \ref{GeometricRV}, Algorithm {\tt GeometricRV} guarantees rendezvous
of agents at some point in time $t$, for the walk $f$ starting at $v$ and the walk $g^*$ starting at $w''$.

Consider the positions at time $t$ of both agents starting at $v$ and $w$, in walks $f$ and $g$.
Since $f(t)=g^*(t)=\Phi ^{-1}(g(t))$, and $\Phi$ is a translation by a vector of length less than $r$ (hence less than $\epsilon$),
it follows that the distance between $f(t)$ and $g(t)$ is less than $\epsilon$. Since walks $f$ and $g$ were arbitrarily
chosen by the adversary, this guarantees $\epsilon$-approximate rendezvous.  
\end{proof}

A consequence of Theorem \ref{approx} is that if agents have arbitrarily small positive visibility ranges
(rather than 0 visibility range as we assumed) and they start in arbitrary points of the (empty) plane,
then Algorithm {\tt GeometricRV} guarantees that
they will see each other in finite time, 
regardless of the actions of the adversary. 

\section{Conclusion}

We provided deterministic asynchronous rendezvous algorithms for graphs and for terrains in the plane.
We studied only the feasibility of rendezvous and
our results are very general: for the graph scenario, we showed that
rendezvous is possible in any connected countable (finite or infinite) graph, starting from any nodes, without any information
on the graph. The only thing an agent needs to know is its own label. In particular, this result
implies a positive solution of a problem from~\cite{DGKKPV}.

Our algorithms rely on an arbitrary fixed enumeration of quadruples $(i,j,s',s'')$, where $i$ and $j$ are positive integers
and $s'$ and $s''$ are finite sequences of positive integers. The complexity of the algorithm (measured by the worst-case
length of paths that the agents have to traverse until rendezvous) depends on this 
enumeration, but we do not think any enumeration can make the algorithm efficient. Thus a natural interesting question left 
for further investigations is the following:
\begin{quotation}
Does there exist a deterministic asynchronous rendezvous algorithm, working for all connected countable unknown graphs,
with complexity polynomial in the labels of the agents and in the initial distance between them?
\end{quotation}

%%%%%%%%%%%%%%%%%%%%%%%%%%%%%%%%%%%%%%%%%%%%%%%%%%%%%%%%%%%

%%%%%%%%%%%%%%%%%%%%%%%%%%%%%%%%%%%%%%%%%%%%%%%%%%%%%%%%%%% 

\end{document}